*Article*

# Ability-Based Methods for Personalized Keyboard Generation

Claire L. Mitchell [1,2], Gabriel J. Cler [3], Susan K. Fager [4], Paola Contessa [1,2], Serge H. Roy [1,2], Gianluca De Luca [1,2], Joshua C. Kline [1,2] and Jennifer M. Vojtech [1,2,*]

[1] Delsys, Inc., Natick, MA 01760, USA; cmitchell@delsys.com (C.L.M.); pcontessa@delsys.com (P.C.); sroy@delsys.com (S.H.R.); gdeluca@delsys.com (G.D.L.); jkline@delsys.com (J.C.K.)
[2] Altec, Inc., Natick, MA 01760, USA
[3] Department of Speech and Hearing Sciences, University of Washington, Seattle, WA 98105, USA; gcler@uw.edu
[4] Institute of Rehabilitation Science and Engineering, Madonna Rehabilitation Hospital, Lincoln, NE 68506, USA; sfager@madonna.org
* Correspondence: jvojtech@delsys.com

**Abstract:** This study introduces an ability-based method for personalized keyboard generation, wherein an individual's own movement and human–computer interaction data are used to automatically compute a personalized virtual keyboard layout. Our approach integrates a multidirectional point-select task to characterize cursor control over time, distance, and direction. The characterization is automatically employed to develop a computationally efficient keyboard layout that prioritizes each user's movement abilities through capturing directional constraints and preferences. We evaluated our approach in a study involving 16 participants using inertial sensing and facial electromyography as an access method, resulting in significantly increased communication rates using the personalized keyboard (52.0 bits/min) when compared to a generically optimized keyboard (47.9 bits/min). Our results demonstrate the ability to effectively characterize an individual's movement abilities to design a personalized keyboard for improved communication. This work underscores the importance of integrating a user's motor abilities when designing virtual interfaces.

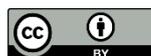

## 1. Introduction

In recent years, mobile devices have increasingly begun to integrate customizable features to provide users with the experience of a more directed, or "personalized," service. Far more than added convenience, these customizations often provide valuable assistance for people with complex communication needs who rely on personal computers, tablets, or smartphones to supplement or replace their oral speech [1–3]. Accessibility features such as increased text size and weight, reachability, magnification, and color inversion are among the myriad of customization options that allow people who rely on augmentative and alternative communication (AAC) to successfully engage with others [4].

*1.1. Motivation*

Although there exists a wide range of customizable device features to facilitate AAC, some individuals remain poorly served. Those with concomitant motor impairments—as in some developmental disabilities (e.g., cerebral palsy, muscular dystrophy), acquired neurogenic disorders (e.g., stroke, traumatic brain injury), and degenerative neurological conditions (e.g., amyotrophic lateral sclerosis, multiple sclerosis)—often lack the manual dexterity needed to control AAC technology and, as a result, require alternative access through switch scanning, head- or eye-tracking, and/or touchscreen devices accessed via finger or typing stick, among other methods. Unfortunately, current AAC technologies with computer interfaces offer limited access efficiency and subpar personalization options, which do little to facilitate communication for those with severe motor impairments who rely on alternative access methods. This is because customizing these alternative access methods largely fails to fully address or compensate for the access barriers that individuals with severe motor impairments often face. Poor technical knowledge amongst caregivers and support staff (e.g., due to limited training [5]) and precise mounting requirements that necessitate manual device adjustments [6] are among the largest barriers to communication, despite these individuals being matched to a device that best fits their residual motor capabilities and use preferences [7,8]. Because of this, AAC users that require both alternative access and augmentative speech options are often excluded from successful use of AAC devices [9–11], contributing to the nearly one-third of patients who abandon their clinically prescribed AAC device in favor of less-effective dysarthric speech or manual gestures, among other communication methods [12,13]. In addition to customizing an individual's computer access method, another strategy to achieve more efficient communication when using an alternative access strategy could be to personalize the individual's computer interface.

Current computer interface customization methods to improve AAC devices beyond lexical prediction and/or manually personalizing keyboard content [9] focus on computational strategies for universally rearranging elements within an AAC interface. These strategies often aim to decrease the time needed to traverse a keyboard (such as the ubiquitous QWERTY keyboard) by organizing frequently occurring characters closer to each other. One popular method leverages Fitts-Digraph Energy, a cost function that weights the travel time between keys by the frequency of character-to-character ("digraph") transitions within a language. Fitts-Digraph Energy is commonly used for solving the Metropolis algorithm in order to reduce the probability of accepting subpar keyboard layouts [14–16]. This method has been examined for single-input keyboard optimization to create a layout of alphabetic [16,17] and phonemic [18] keys (the latter of which validated Fitts-Digraph Energy—and, by proxy, Fitts' Law—for examining motor performance in individuals with motor impairments). Although these computational approaches are an *automated* option for increasing communication rates [14–16] through more efficient virtual keyboard layouts, they do not account for any element of *personalization* that is often critical for AAC users who rely on alternative computer access.

*1.2. Ability-Based Design for AAC*

With this absence of AAC technology that *automatically personalizes* to the individual, the field of accessible computing has recently focused on designing devices directly around an individual rather than training an individual to use a piece of technology out of the box. This technique, known as *ability-based design* [19,20], is based on the principle of designing technology that conforms to an individual as opposed to an individual conforming to the technology. An example of this is the SUPPLE system [21], which comprises automated methods for designing a user interface based on the dexterity of an individual's control and their preference for specific graphical elements.

Methods that utilize ability-based design for keyboard optimization are substantially limited for AAC use. Recent work by Sarcar and colleagues [22] merged adjacent keys from the QWERTY keyboard and adapted parameters—such as key size and number of predicted words—to specific diseases, including tremor, dyslexia, and memory dysfunction. While the authors present methods that potentially improve text-entry speed, this work has significant limitations in that (i) the complex and heterogenous motor and communication needs of AAC users even within a single disease population make it difficult to effectively generalize keyboard interactions by disease [23,24]; (ii) ability-based optimization methods were validated in only two participants; and (iii) the authors adopt the inefficient, multi-input QWERTY layout for use with a single-input access method [14,15,25–29]. Methods that account for user-specific movement patterns have been developed for automatically adapting key presses on a touchscreen [30], wherein adapted key-press classification models discriminate between actual and intended key presses. Unfortunately, this method is only validated for multi-input typing using a QWERTY layout. There remains a need for ability-based optimization of *single-input keyboards* for use by people with complex communication needs and concomitant motor impairments.

Applying the principles of ability-based design to the field of AAC is an important step toward developing technology that can effectively serve any individual who relies on alternative communication methods. As of today, personalized AAC technology is still largely limited to case-by-case examples of manual customization due to the immensely complex and heterogenous motor function of this population [31]. Thus, there is a crucial need for AAC technology that can be *automatically* and *directly tailored* to prospective users.

*1.3. Current Investigation*

Using principles of ability-based design, the purpose of this work was to overcome limitations with current single-input AAC interface technology to provide a solution that automatically arranges a keyboard interface to an individual's cursor control or residual motor ability to improve communication efficiency. To achieve this goal, we developed an AAC system that calibrates a single-input access method to an individual's motor abilities, and then uses those same motor abilities to automatically organize a personalized keyboard for the individual.

Our approach integrates established principles of Fitts-Digraph Energy following the work of Cler and colleagues (2019) in optimizing a phonemic keyboard for individuals with motor impairments to computationally optimize a keyboard of orthographic characters (i.e., English letters A–Z, space) to an individual [18]. However, whereas Fitts-Digraph Energy is classically computed using a single set of generic movement constants to describe anticipated user movement while traversing an interface (e.g., [14–16,18,25,26,29,32,33]), in this study we examine the feasibility of using person- *and* direction-specific constants within Fitts' Law to characterize movement for optimizing a directionally personalized keyboard. Movement direction is an important variable in ability-based design since AAC users may have conditions that preclude access to the full interface—such as limited or unequal range of movement (e.g., due to cerebral palsy) or a visual field cut or condition that results in peripheral focus (e.g., due to brainstem stroke) [34,35]—and because directional performance differences have been observed even in individuals without motor impairments [36–39]. These preferences and abilities could result in, for example, keyboards that are more vertically oriented for individuals that have difficulty moving a cursor left to right, or cross-shaped keyboards for individuals that have difficulty controlling diagonal cursor movement or prefer to move the mouse in orthogonal $x$ or $y$ directions at a time.

As such, we tested our personalization methods among 16 participants without motor impairments when using an alternative access method to evaluate method effectiveness and determine feasibility for testing in the anticipated target population of AAC users who require alternative access. Although the motor control of AAC users has been specifically observed to differ with intended movement direction [34,35] (thus presumably resulting in a personalized keyboard reflecting those abilities), motor control has been shown to substantially differ across movement direction even in individuals without motor impairments [36–39]. We argue that these feasibility results lay the groundwork for the development of personalized keyboards for individuals with constrained and/or uneven mobility. Communication performance was examined while using this access method to create messages using a series of two-dimensional (2D) interfaces: (i) a keyboard generated using our personalization methods, (ii) a keyboard generated via traditional optimization parameters, and (iii) the ubiquitous QWERTY keyboard.

We hypothesized that participants would exhibit movement strategies that differed throughout the possible range of motion within a 2D virtual interface using an alternative access method [40]. Due to hypothesized differences with respect to direction, we further hypothesized that the personalized keyboards that integrated a user's directional information would lead to better communication performance when compared to both a generically optimized keyboard, as well as a QWERTY keyboard.

**2. Materials and Methods**

In this section, we detail the methods used to generate virtual keyboards, followed by a description of the experimental study used to evaluate the keyboards.

*2.1. Keyboard Personalization*

We designed algorithms to characterize an individual's 2D cursor control abilities by capturing unique relationships between movement time and distance over different directions of movement. These algorithms leverage an expanded version of Fitts' Law to estimate 2D movement time and distance relationships relative to a given target angle (i.e., rather than the typical approach of grouping time and distance data irrespective of angle) within a modified multidirectional point-select task, as described in detail below.

2.1.1. Movement Characterization

The goal of the task was to navigate to and select the highlighted targets amongst a screen of blank keyboard keys configured in a 2D grid to capture movement control data across a range of movement distances and directions. Specifically, we arranged hexagonal keys in a 9 × 9 honeycomb grid. Sequential targets were presented across 16 angular ranges with respect to one another: four to capture the cardinal directions provided by the access method (i.e., up, down, left, right), four for the intercardinal directions that bisect the four cardinal directions, and eight to capture the half-directions that bisect the cardinal and intercardinal directions (see Figure 1a and Figure 1b).

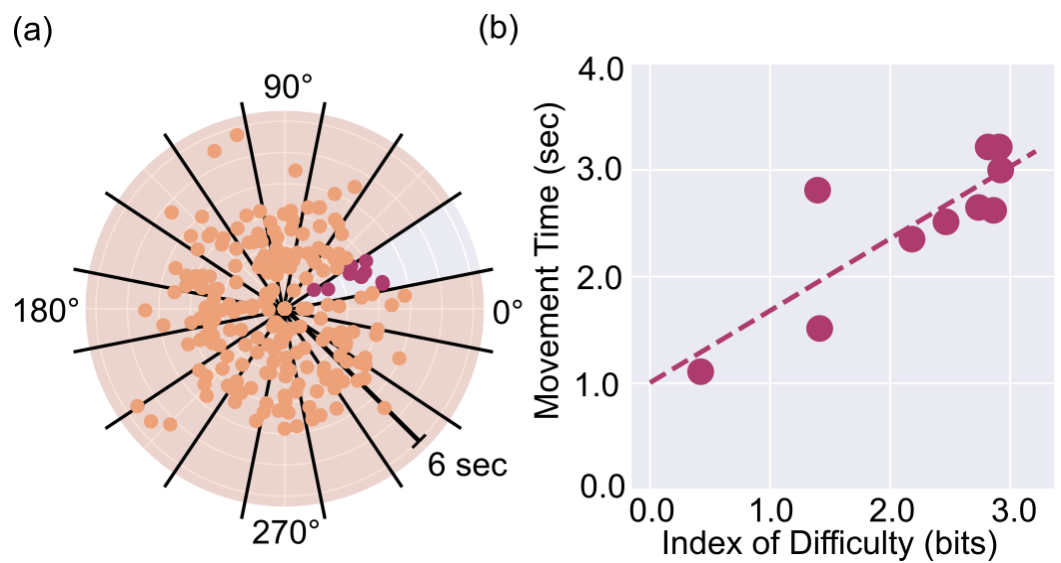

**Figure 1.** Example target selection data from movement characterization task. (**a**) Movement time shown as a function of target selection angle, split into 16 angular ranges (black lines). One section is highlighted with pink dots to highlight data collected within a single angular range. (**b**) Example regression of movement time against index of difficulty (bits) from the highlighted data taken in (**a**) to obtain Fitts' Law constants.

The task was initially seeded with 225 targets to ensure that at least 10 movement trajectories of varying distance could be captured for each angular range [41]. Distance (D) was calculated as the Euclidean distance between sequential target click locations. Using Mackenzie's Shannon formulation of Fitts' Law (Equation (1); [42,43]), each distance

*D* was converted to an index of difficulty (*ID*) with *W* representing a constant target width (pixels). Actual target width *W* was used instead of effective target width, *We*, since—given unlimited chances to choose a target—the error rate is 0%. This also avoids misestimation of bits/s classically associated with *W* as opposed to *We* [44]. Movement time (*MT*) to select the target was calculated as the travel time of the cursor between clicks on sequential targets.

$$MT = a + b \times ID, \text{ where } ID = \log_2\left(\frac{D}{W} + 1\right) \quad (1)$$

Nine unique *IDs* were used ranging from 0, the minimum inter-key distance to capture an estimate of user movement (*a* in Equation (1)), to 8, the maximum inter-key distance in one direction. Unlimited chances to select a target were provided to each participant to effectively capture accidental clicks as well as repeatable click errors relative to a given movement distance and/or direction. Twenty-five targets were seeded for each *ID*. The resulting sequence of 225 targets were randomized and organized into a queue.

At completion of the initial target selection task, linear regressions were performed across each angular range tested to derive angle-specific regression constants a and b of Fitts' Law (Equation (1)). To account for the possibility that occasional target selection errors led to outlier movement times, the target selection process was repeated for any angular ranges with weak *MT-ID* correlations (via a coefficient of determination, or $R^2$, ≤0.25 [45]) to glean more representative distance–direction information for the participant. This process was repeated until either the number of targets reached 400—an empirically determined cut-off to maintain a task time approximately under 20 min and to minimize the possibility of participant fatigue—or if each angular range contained 10 or more targets, exhibited a moderate or better correlation between *MT* and *ID* ($R^2 > 0.25$ [45]; Figure 1b), and all outliers (defined as a target ±3 standard deviations from the regression) had been successfully repeated. During the task, participants were able to signal to the experimenter through hand gestures if a break was needed—in such cases, participants were instructed to relax for two minutes and refrain from selecting targets.

2.1.2. Personalized Keyboard Generation

Personalized keyboards were created from the ability-based keyboard optimization algorithm by leveraging digraph transition occurrences (i.e., representing letter-to-letter transitions when spelling messages) and user *MT* relative to both *ID* and target selection angle. As detailed below, these keyboards were designed by solving the quadratic assignment problem (QAP) using the GraphMatch function in the Python *graspologic* library (Microsoft, Redmond, WA, USA; [46,47]) as it does not require hyperparameter tuning (unlike the Metropolis algorithm [48]).

The QAP problem is designed to minimize the cost of arranging *N* items where the cost is proportional to the flow and distance between items [49]. In this application of the QAP problem, the flow between items is represented by all possible digraph transitions between *N* characters (English letters A–Z, space), resulting in a matrix of size *N* × *N*. Distance is represented as movement time between all possible positions M to place a target, thus is an *M* × *M* matrix. Values for this matrix were determined in a three-step process consisting of (i) calculating the target selection angle between every *M* position, (ii) sampling a and b user-specific Fitts' Law constants for the specified angle and (iii) applying resulting constants to Fitts' Law (Equation (1)) to derive *MT*. In this way, unique movement times are provided for a given user. With the flow and distance matrices successfully populated, the personalized keyboard arrangement is then configured via the Fast Approximate QAP Algorithm of the GraphMatch function. In the current study, the digraph transition occurrences for the keyboard personalization algorithms were obtained from a corpus of phrase sets for evaluating text-entry techniques [50]. Each keyboard included 27 keys comprising the 26 English orthographic letters and a space.

## 2.2. Keyboard Evaluation

The personalized keyboard methods were evaluated in a series of experimental sessions amongst 16 participants. The sessions included tasks to characterize user movement and generate a unique, virtual keyboard for a given participant, as well as transcription tasks to evaluate keyboard performance relative to a generically optimized keyboard (i.e., the current state-of-the-art) and the ubiquitous QWERTY (typewriter-style) keyboard. Evaluation methods are described in detail below.

### 2.2.1. Experimental Overview

To quantify any communication benefits presented by our personalized keyboard methods (example keyboard shown in Figure 2a), communication performance was compared against a generically optimized keyboard (Figure 2b) as well as the ubiquitous QWERTY keyboard (Figure 2c) for 16 participants. Whereas personalized keyboard generation included custom *a* and *b* Fitts' Law constants relative to movement direction, the optimized keyboard was generated using standard Fitts' Law constants of cursor movement with delay (*a* = 0.127 s) and acceleration (*b* = 1/4.9 s/bits) across all target selection angles [14,29]. Our QWERTY keyboard was arranged following the standard QWERTY (typewriter-style) layout except for the space key; to be able to compare performance using QWERTY to the other keyboards (i.e., personalized, optimized), the width of the space key was set equal to all other keys and positioned to the right of the "M" key (see Figure 2c).

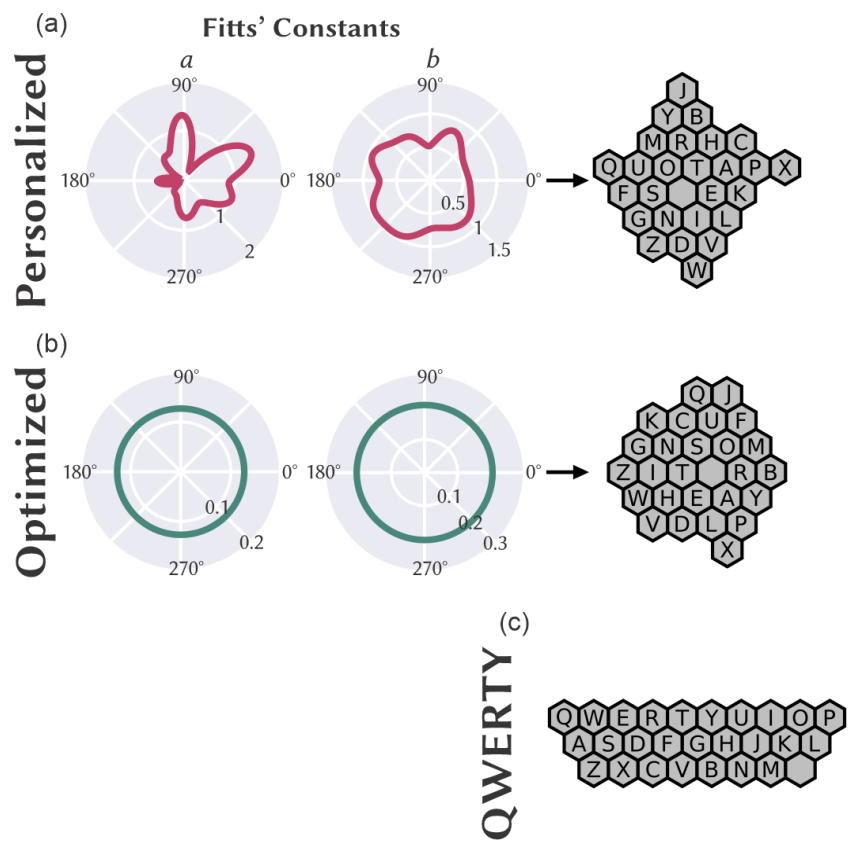

**Figure 2.** Keyboards used for study and their generation from Fitts' Law constants when relevant. (**a**) Example of Fitts' Law constants (**left**, **middle**) and keyboard (**right**) for one participant. (**b**) The static Fitts' Law constants (**left**, **middle**) used to generate the optimized keyboard (**right**). (**c**) The QWERTY keyboard.

The experiment comprised nine sessions, each on unique days, to assess participant communication ability. Prior to the experiment, participants were pseudorandomly assigned a keyboard order between optimized and personalized keyboards: generically optimized first or personalized first. Order assignment was counterbalanced across participants to ensure an equal number followed each order. Performance using QWERTY was analyzed in the final session—i.e., after participants were familiar with the access method—to serve as a reference for communication performance as participants did not need training with QWERTY due to widespread familiarity in using it for mobile device communication (see self-reported familiarity scores with QWERTY in Section 2.2.2).

2.2.2. Participants

Sixteen individuals (8 cisgender female, 8 cisgender male; 27.9 ± 5.1 years) without history of speech or motor impairments participated in the study. All individuals gave written, informed consent in compliance with the Western Institutional Review Board (WIRB Protocol #20192468, approved 23 September 2021). According to self-reports based on Likert scale ratings, all participants were proficient in English (6.7 ± 0.8, where 1 = "Very Bad" and 7 = "Very Good") and familiar with QWERTY (6.3 ± 1.2, where 1 = "Not Familiar" and 7 = "Very Familiar").

2.2.3. Sessions

All participants completed nine experimental sessions, each lasting 1–1.5 h. Sessions with consecutive use of the same keyboard were performed 24 h apart; all other sessions were separated by a minimum of 24 h except for one participant who participated in sessions 1 and 2 separated by a 3-h break. Each session comprised sensor application and calibration, a short familiarization task for participants to test their movement and click control, then either the movement characterization task or keyboard communication task. Within a session, participants were exposed to one of five keyboards: one of two generically optimized keyboards (vertically flipped versions of each other), one of two personalized keyboards (from the movement characterization tasks in sessions 1 and 5), or the QWERTY keyboard.

At the start of the first session, experimenters explained that the purpose of the study was to evaluate a set of keyboards, but did not describe the differences between the keyboards, how they were generated, or the expectations for performance between keyboards. After carrying out the familiarization task, participants completed the first movement characterization task; the resulting movement data were used to inform the design of the first personalized keyboard.

The next four sessions (2–5), referred to as the "first block", required participants to perform the communication task using their first assigned keyboard (optimized or personalized first) for sessions 2 and 3, then their second assigned keyboard for sessions 4 and 5. The movement characterization task was repeated at the end of session 5 to generate a new personalized keyboard. The next four sessions (6–9), called the "second block," required participants to use their second personalized keyboard as well as a flipped version of the optimized keyboard per their assigned keyboard order. Instead of repeating the movement characterization at the end of session 9, participants took a break to minimize fatigue and then carried out the keyboard evaluation task using QWERTY. An overview of these experimental sessions is shown in Figure 3.

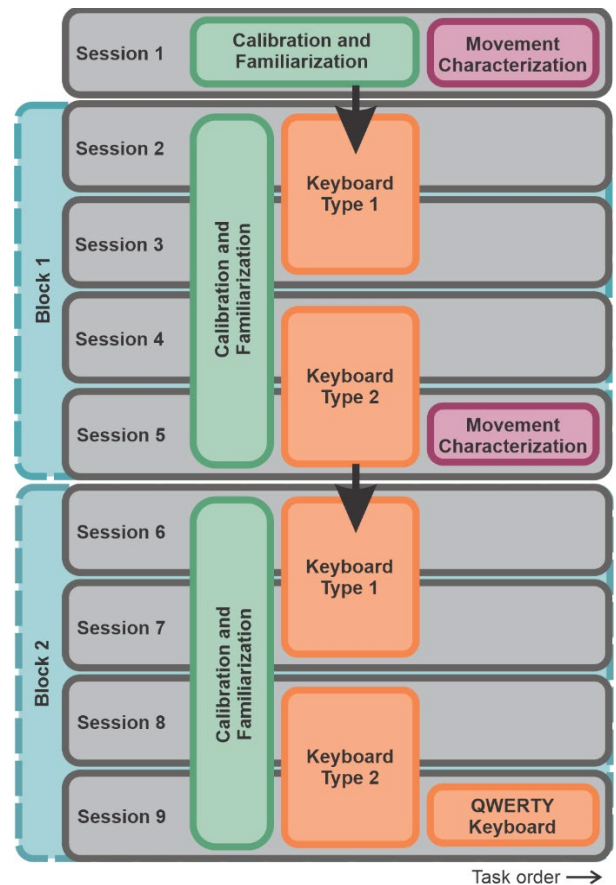

**Figure 3.** Organization of the 9-session study.

2.2.4. Access Method Setup and Calibration

Computer access in this study was provided via a method that combines surface electromyographic (sEMG) sensing of musculature and inertial sensing (IMU) of motor capabilities to control cursor clicks and cursor movement, respectively. The hybrid sEMG/IMU access method demonstrates flexibility across environmental conditions [40,51], showing promise for single-input cursor control for those with severe motor impairments who cannot maintain one body position (e.g., due to changes in posture by caregivers or the users themselves) and require continuous use throughout the course of the day. Hybrid sEMG/IMU access was provided through a single Trigno Mini sensor (Delsys, Natick, MA, USA). IMU signals were sampled at 148 Hz and comprised tri-axial ($x$, $y$, $z$) acceleration signals, whereas sEMG signals were sampled at 2222 Hz, band-pass filtered between 20 and 450 Hz, and amplified by a gain of 300. Signals were transmitted wirelessly from the sEMG/IMU sensor to a Trigno acquisition system and digitally acquired via a custom Delsys API Python wrapper.

Using methodology from [40], the sEMG/IMU access method was configured to translate the gravitational acceleration vector from tri-axial acceleration signals into tilt angles that correspond to the velocity of the cursor movement. To do so, acceleration signals were first averaged over 54 ms windows, then converted to tilt angles ($\beta$, $\gamma$, $\theta$) from rectangular coordinates ($x$, $y$, $z$). Incoming tilt angle signals were detrended and normalized via values as calculated during system calibration. The resulting control was thus specific to the range of head tilt angles exhibited by the individual, with smaller tilt angles corresponding to lower velocity cursor movements while maintaining 360-degree control. To perform clicks, the root-mean-square (RMS) values obtained over 54 ms windows of the sEMG signal were calculated and a click occurred if the RMS value exceeded 70% of the maximum RMS set during the calibration. Once a click was activated, a second click could not be performed until the RMS fell below 30% of the maximum RMS to minimize double-clicks. These thresholds for determining clicking behavior were adopted from [52].

At the beginning of each session, the skin surface was prepared by cleaning the sensor sites with alcohol pads and gently exfoliating with medical-grade adhesive tape to remove excess dead skin and oils [53–55]. Double-sided medical adhesive tape was used to secure the body of the sEMG/IMU sensor to the center of the forehead, with the *y*-axis of the IMU parallel to the transverse axes of the head and the EMG sensor component applied over the orbicularis oculi of the preferred eye ($N$ = 7 left, $N$ = 9 right).

Computer access thresholds were calibrated by instructing each participant to comfortably tilt their head to the left, right, up, and down twice, and wink or hard blink twice [40,51]. These data were used to tune the 2D range of cursor movement from head tilt angle (left, right, up, down) and threshold for cursor clicks from eyeblink activity. Participants tested each calibration by navigating in different directions on the interface and selecting multiple targets. Calibrations were repeated if participants were not satisfied with their control (e.g., poor or inconsistent movement and/or click control).

2.2.5. Virtual Interface Setup and Evaluation

An external monitor of resolution 1920 × 1080 pixels was used to display the virtual interfaces and was connected to a laptop controlled by the experimenters. Experimental software presented to participants was built in Python 3, relying on the following open-source packages: pandas [56,57], SciPy [58], NumPy [59], Matplotlib [60], graspologic [46], and PyInstaller [61]. The software displayed the assigned interface to participants (movement characterization task or personalized/optimized/QWERTY keyboard) on a gray screen with hexagonal keys. All key shapes were hexagonal, as this shape has been shown to allow effective movement between keys [16,48]. All keys were set to a width of 130 pixels to contain up to 9 keys horizontally or vertically to accommodate keyboards of different shapes and orientations given the monitor resolution. Audible click feedback was played through speakers when participants clicked using the sEMG/IMU access method. An example of an individual using the setup is shown in Figure 4.

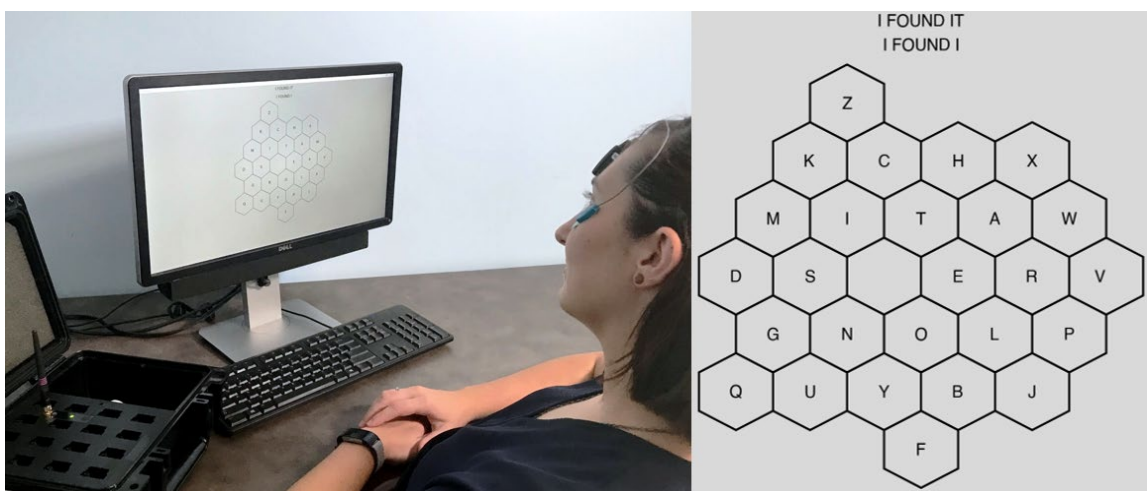

**Figure 4.** Example experimental setup and personalized keyboard. An individual navigates their personalized keyboard to spell "I FOUND IT" using the sEMG/IMU access method.

Sessions involving the movement characterization task were carried out within the 9 × 9 honeycomb grid described in Section 2.1.1. The three keyboard types (personalized, generically optimized, QWERTY) were evaluated in a separate communication task in which participants used the sEMG/IMU access method to navigate to and select keys to spell out a set of prompts presented from a corpus for evaluating text-entry techniques [50]. The communication task was self-paced, wherein participants first pressed "Enter" on a physical keyboard when they were ready to begin a trial. After pressing "Enter," a prompt would appear above the virtual keyboard and the letters on the keyboard would disappear; this was done so the participants would focus on the words in the prompt as opposed to plotting the path they would take on the keyboard (see Figure 3). Once ready to begin spelling out the prompt using the virtual keyboard, participants pressed "Enter" a second time on the physical keyboard to make the letters on the virtual keyboard reappear. At this point, participants spelled out the prompt, pressing "Enter" a final time to end the trial once they had finished selecting characters. Participants were instructed to continue without interruption if mistakes were made during the spelling process [18]. Participants took breaks as needed between trials and repeated this process for a series of 20 prompts within a given keyboard.

*2.3. Data Analysis*

Primary outcome measures for each evaluation aimed to capture both participant-specific movement characteristics as well as each participant's communication ability. To capture movement characteristics, we measured target selection accuracy (%), speed via words per min (*WPM*), and information transfer rate (*ITR*; bits/min). *WPM* was assessed as the number of characters, correct or incorrect, per minute divided by an average of 5 characters per word [16]. Accuracy was either 100% when the target was successfully selected or 0% when a participant failed to select it. *ITR* was measured using Wolpaw's method to consider *MT* and accuracy relative to the number of possible targets [62].

Statistical analysis was performed in jamovi (version 1.8; [63–66]). A series of linear mixed-effects models (LMMs) were constructed to examine the effect of the keyboards on outcome measures of accuracy, WPM, and ITR when considering the random effects of the participant. Target selection accuracy data were transformed prior to parametric testing via a Box-Cox transformation to account for deviations from normality.

A set of LMMs were first implemented to understand the effects of participant (random), computational keyboard efficiency (covariate)—calculated using Fitts-Digraph Energy [16]—and fixed effects of keyboard (personalized, optimized), keyboard exposure (first exposure to keyboard, second exposure to keyboard), as well as keyboard block (sessions 2–5, sessions 6–9) and the interactions of keyboard × exposure and keyboard × block on each of the outcome measures (i.e., target selection accuracy, *WPM*, *ITR*). These fixed effects were treated as within-subject factors. Computational keyboard efficiency was included here as a covariate since two different personalized keyboards were implemented, which may not only stem from variability in the participant, but also from differences in motor ability as individuals learn to use the sEMG/IMU access method.

An additional set of LMMs were then constructed to examine communication performance between the second personalized keyboard (used within the second keyboard block) and QWERTY (fixed factor) when considering the random effects of participant. For this analysis, keyboard (personalized, QWERTY) was treated as a within-subject factor. Given the nontraditional placement and size of the space key in our QWERTY keyboard (Figure 2c) relative to the standard QWERTY keyboard, we compared personalized and QWERTY keyboards with and without movements to and from the space key on each keyboard. The metric that disregarded involving the space key (*WPM\**) was included to ensure that *WPM* for QWERTY would not be skewed lower from our unique placement of the space key alone.

For each set of LMMs, an $\alpha$ level of 0.05 was used. Effect sizes were estimated for fixed factors using partial eta squared ($\eta_p^2$), interpreted with cutoffs of 0.01, 0.06, 0.14 for small, medium, and large effect sizes, respectively [45]. Post hoc analyses were conducted on significant main effects via Tukey simultaneous tests using a Bonferroni correction for multiple comparisons.

**3. Results**

The findings for the movement characterization and keyboard evaluation tasks are described for the 16 participants below.

### 3.1. Movement Characterization

The participants were presented with an average of 283.4 targets (*SD* = 46.8) to characterize their movement patterns. The linear regressions performed on the resulting movement data to compare Fitts' Law-based *MT* to target *ID* exhibited an average $R^2$ = 0.55 (*SD* = 0.02, *range* = 0.06–0.96). Additionally, Fitts' Law constants computed relative to movement direction produced average values of *a* = 0.83 (*SD* = 0.53) and *b* = 0.91 (*SD* = 0.44) and can be seen in more detail for session 5 in Figure 5.

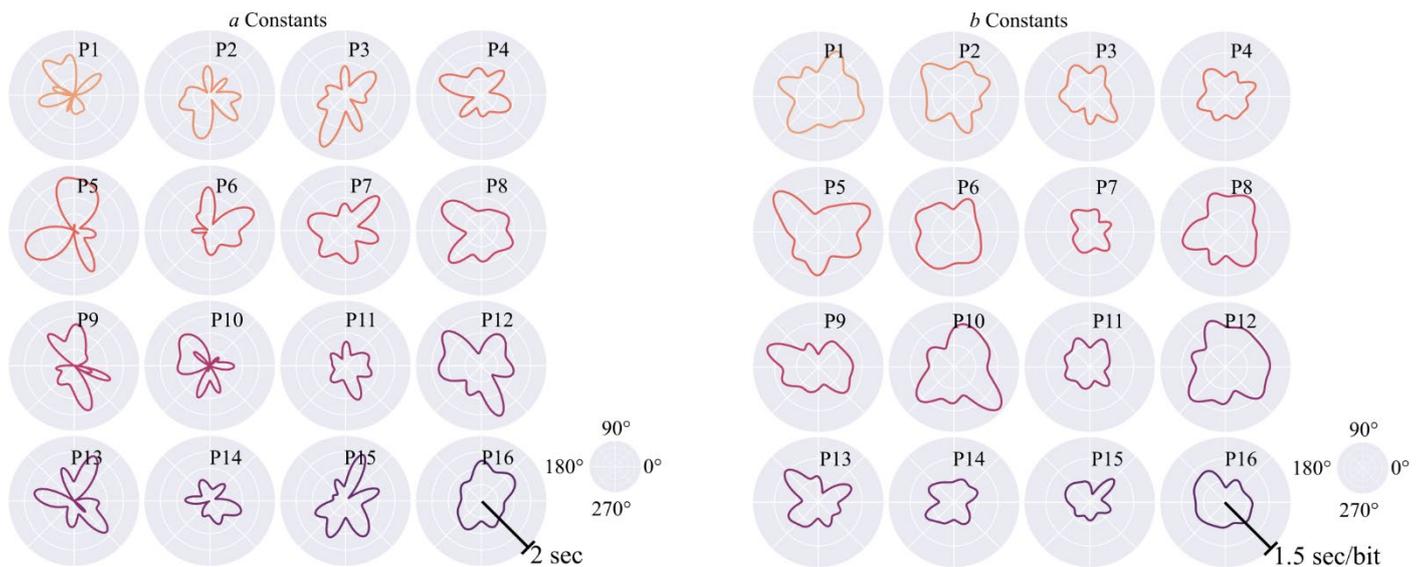

**Figure 5.** Fitts' Law constants a and b for each participant. Constants *a* (**left**) and *b* (**right**) interpolated throughout the range of motion for participants (e.g., P1) for session 5.

### 3.2. Optimized vs. Personalized Keyboards

Table 1 displays the model summaries constructed for target selection accuracy, WPM, and ITR between optimized and personalized keyboards. For all models, no significant interaction effects (keyboard × block, keyboard × exposure) were observed ($p \geq 0.05$).

**Table 1.** Linear mixed-effects models for outcome measures comparing generically optimized and personalized keyboards.

| Model | Effect | *df* | $\eta_p^2$ | *F* | *p* |
|---|---|---|---|---|---|
| Target Selection Accuracy | Efficiency | (1, 129) | – | 2.75 | 0.100 |
| | Keyboard | (1, 74) | – | 1.79 | 0.185 |
| | **Block** | **(1, 2519)** | **0.01** | **15.07** | **<0.001** |
| | **Exposure** | **(1, 2518)** | **0.00** | **7.55** | **0.006** |
| | Keyboard × Block | (1, 2519) | – | 1.72 | 0.190 |
| | Keyboard × Exposure | (1, 2518) | – | 0.27 | 0.605 |
| WPM | **Efficiency** | **(1, 2473)** | **0.02** | **43.33** | **<0.001** |
| | **Keyboard** | **(1, 24)** | **0.23** | **6.93** | **0.015** |
| | **Block** | **(1, 2519)** | **0.29** | **1045.90** | **<0.001** |
| | **Exposure** | **(1, 2518)** | **0.09** | **249.95** | **<0.001** |
| | Keyboard × Block | (1, 2519) | – | 0.00 | 0.971 |
| | Keyboard × Exposure | (1, 2518) | – | 1.30 | 0.254 |
| ITR | **Efficiency** | **(1, 2403)** | **0.01** | **33.24** | **<0.001** |

| Model | Effect | df | $\eta_p^2$ | F | p |
|---|---|---|---|---|---|
| | **Keyboard** | **(1, 27)** | **0.21** | **6.87** | **0.014** |
| | **Block** | **(1, 2519)** | **0.25** | **845.75** | **<0.001** |
| | **Exposure** | **(1, 2518)** | **0.07** | **188.21** | **<0.001** |
| | Keyboard × Block | (1, 2519) | – | 0.16 | 0.692 |
| | Keyboard × Exposure | (1, 2518) | – | 0.46 | 0.482 |

Note. *df* = degrees of freedom (numerator, denominator), $\eta_p^2$ = partial eta squared, *F* = *F*-statistic, *p* = *p*-value, WPM = words per minute, ITR = information transfer rate. Dashes (–) indicate non-significant findings ($p \geq 0.05$). **Bold** rows indicate significant effects.

### 3.2.1. Target Selection Accuracy

The model for accuracy showed significant effects for block ($p < 0.001$) and exposure ($p = 0.006$); however, post hoc analyses of the fixed main effects showed that effect sizes were small for block ($\eta_p^2 = 0.01$) and negligible for exposure ($\eta_p^2 < 0.01$). No significant effects were seen for efficiency or keyboard (Figure 6a).

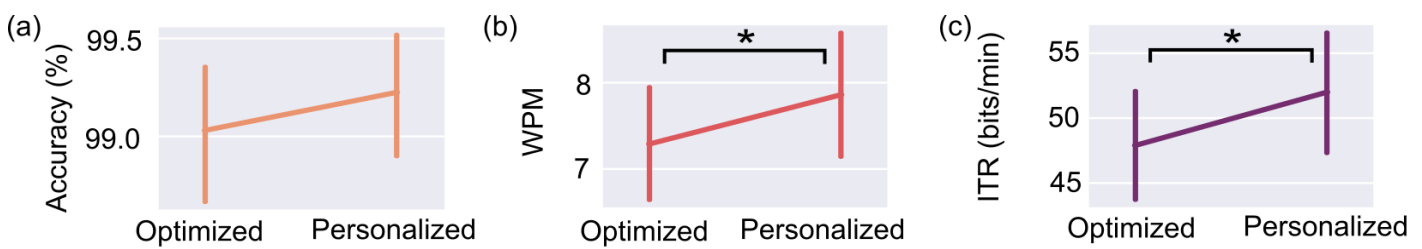

**Figure 6.** Marginal means for outcomes of transcription task when comparing optimized and personalized keyboards. (**a**) Target selection accuracy, (**b**) words per minute (*WPM*), and (**c**) information transfer rate (*ITR*) shown across keyboards. Error bars represent 95% confidence intervals. \**p* < 0.05.

### 3.2.2. WPM

The model for *WPM* revealed a significant, small effect for computational keyboard efficiency ($p < 0.001$, $\eta_p^2 = 0.02$); a significant, medium effect for exposure ($p < 0.001$, $\eta_p^2 = 0.09$); and significant, large effects for both keyboard ($p = 0.015$, $\eta_p^2 = 0.23$) and block ($p < 0.001$, $\eta_p^2 = 0.29$). Post hoc analyses of the fixed main effects revealed that personalized keyboards averaged 0.57 wpm greater than the optimized keyboards (7.86 wpm vs. 7.29 wpm; see Figure 6b). Higher *WPM* were also observed when participants were exposed to a keyboard for the second time (7.83 wpm vs. 7.32 wpm) and when carrying out the second experimental block (8.11 wpm vs. 7.05 wpm).

### 3.2.3. ITR

The model for *ITR* showed a significant, small effect of computational keyboard efficiency ($p < 0.001$, $\eta_p^2 = 0.01$); significant, medium effect for exposure ($p < 0.001$, $\eta_p^2 = 0.07$); and significant, large effects for both keyboard ($p = 0.014$, $\eta_p^2 = 0.21$) and block ($p < 0.001$, $\eta_p^2 = 0.25$). Post hoc analyses of the fixed main effects revealed higher average *ITR* values when participants used their personalized keyboards (52.0 bits/min) compared to the optimized keyboards (47.9 bits/min; see Figure 6c). Higher average *ITR* values were also observed when participants were exposed to a keyboard for the second time (51.7 bits/min vs. 48.1 bits/min), as well as during the second experimental block compared with the first (53.8 bits/min vs. 46.1 bits/min).

### 3.3. Personalized vs. QWERTY Keyboards

Target selection accuracy data were first transformed via a Box-Cox transformation to meet the assumptions of normality for the planned parametric LMM. The resulting LMMs showed that the keyboard (personalized, QWERTY) exhibited a significant, large main effect in the models for target selection accuracy ($p = 0.005$, $\eta_p^2 = 0.42$), WPM ($p < 0.001$, $\eta_p^2 = 0.51$), WPM* ($p < 0.001$, $\eta_p^2 = 0.20$), and *ITR* ($p < 0.001$, $\eta_p^2 = 0.87$; Table 2).

**Table 2.** Linear mixed-effects models for outcome measures comparing personalized and QWERTY keyboards.

| Model | Effect | df | $\eta_p^2$ | F | p |
| --- | --- | --- | --- | --- | --- |
| Target Selection Accuracy | Keyboard | (1, 15) | 0.42 | 10.8 | 0.005 |
| WPM | Keyboard | (1, 15) | 0.51 | 71.6 | <0.001 |
| WPM* | Keyboard | (1, 15) | 0.20 | 74.5 | <0.001 |
| ITR | Keyboard | (1, 15) | 0.87 | 97.2 | <0.001 |

Note. *df* = degrees of freedom (numerator, denominator), $\eta_p^2$ = partial eta squared, *F* = *F*-statistic, *p* = *p*-value, WPM = words per minute, WPM* = words per minute without space, ITR = information transfer rate.

Post hoc analyses revealed that communicating using QWERTY led to significantly lower mean target selection accuracies than when using personalized keyboards (99.0% vs. 99.4%; Figure 7a), although all accuracies were very high. Similarly, *WPM* and *WPM** were significantly smaller with QWERTY (6.79 and 7.03 wpm, respectively) compared to personalized keyboards (8.36 and 7.90 wpm, respectively; Figure 7b,c). Personalized keyboards also led to significantly greater average *ITRs* (55.6 bits/min) than QWERTY (44.4 bits/min; Figure 7d).

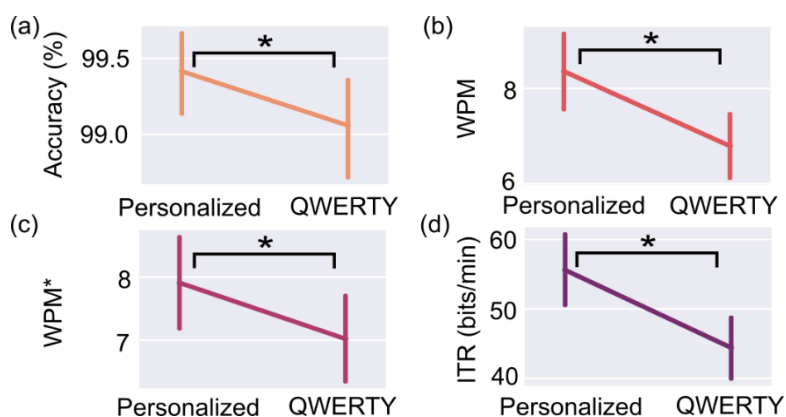

**Figure 7.** Marginal means for outcome measures of transcription tasks comparing personalized and QWERTY keyboards. (**a**) Target selection accuracy, (**b**) words per minute (*WPM*), (**c**) words per minute without space (*WPM**), and (**d**) information transfer rate (*ITR*) shown across keyboards. Error bars represent 95% confidence intervals. * $p < 0.05$.

## 4. Discussion

In this study, methods for automatically configuring a keyboard to an individual's 2D cursor control were built and evaluated amongst participants without motor impairments. This study establishes the feasibility of personalizing a keyboard to an individual through uniquely capturing an individual's preferred movements. Overall, our results support our hypotheses that (i) participants would exhibit diversity in movement capabilities across direction, and (ii) participants' personalized keyboards would lead to greater communication performance when compared to generically optimized and QWERTY keyboards. These findings are described in detail below.

*4.1. Movement Characterization*

In characterizing an individual's cursor control within a 2D interface, we hypothesized that participants would exhibit movement strategies that differed throughout the possible 360 degrees of motion. Indeed, our results support this hypothesis, wherein participants exhibited variability across Fitts' Law constants (*a* and *b* in Equation (1)) for different movement directions (see Figure 5). Our approach to characterizing cursor control via expanding Fitts' Law to encapsulate a table of constants across movement direction, on average, demonstrated a robust ability to capture this variability ($R^2 = 0.55$) with a range of performance spanning small ($R^2 = 0.06$) to large ($R^2 = 0.96$) relationships between expected *MT* and task *ID*. Relationships below the cutoff set in the multidirectional point-select task ($R^2 = 0.25$) were observed when participants reached the 400 target limit; these observations were rare, only occurring in 3.9% of *MT-ID* relationships.

Our results highlight the immense variability in Fitts' Law parameters obtained within and across participants, with standard deviations of 0.53 s for *a* and 0.44 s/bits for *b*. The variability observed in directional Fitts' Law constants for these participants supports the notion that individuals exhibit different control strategies when using the sEMG/IMU access method to navigate a 2D interface and, furthermore, suggests that movement direction may be an important factor to consider when capturing these movements. On average, *a* and *b* Fitts' Law constants were greater (*a* = 0.83 s and *b* = 0.91 s/bits) than those reported in literature (*a* = 0 or 0.127 s and *b* = 1/4.9 s/bits; e.g., [14,29]); however, these findings likely reflect differences in access method compared to those examined in prior works (i.e., sEMG/IMU vs. stylus and touchscreen).

Much of the work using Fitts' Law to assess 2D tasks has been pioneered by MacKenzie [67–73], who has demonstrated the utility of using this "model by analogy" to empirically evaluate user control with a 1D or 2D task. Many studies leveraging MacKenzie's Shannon formulation of Fitts' Law to evaluate movement systematically vary target angle to effectively "smooth over" the effects of movement direction [44,74–76]. To do so, these works typically utilize an ISO-standard multidirectional point-select task that leverages a cluster of circular targets of equal diameter positioned equidistantly around a large circle; by instructing participants to navigate and select diametrically opposite targets, the effects of movement direction can be effectively ignored when relating movement time to task index of difficulty (i.e., via achieving robust $R^2$ values across direction).

Because the current study aimed to design a keyboard interface in which users would not necessarily be choosing diametrically opposite characters to spell messages, we chose not to employ this classic point-select task. Other studies that specifically focus on the effects of movement direction within 2D point-select tasks do indeed provide evidence that the angle of target selection influences performance [36–39]. Thus, we designed our movement characterization task to facilitate user movement in a variety of distances and directions to be able to capture a spectrum of cursor control data. Our findings suggest that there may be some utility to capturing directional variations when using our sEMG/IMU access method for our specific purpose of personalizing a virtual keyboard interface to an individual's unique motor abilities. Future work may find value in exploring other applications for a directionally expanded Fitts' Law, such as for website design or home screen configuration on mobile devices.

*4.2. Keyboard Communication*

The communication rates of the participants across the three keyboard types (optimized, personalized, QWERTY) were well within the range of those described in the literature for similar head tracking-based access methods. Specifically, average *ITR* values have been shown to range from 5.4 to 120.7 bits/min when used by controls for AAC [52,77–79]. Average *ITR* values were similar across keyboards, ranging from 44.4 bits/min when using QWERTY to 47.9 bits/min and 52.0 bits/min, on average, when using optimized and personalized keyboards, respectively. Our average speed, measured through *WPM*, exceeded those presented in the literature for sEMG-based access methods [52,77–79], which reached rates up to 5.8 wpm.

4.2.1. Optimized vs. Personalized Keyboards

We hypothesized that a personalized keyboard would lead to better communication performance compared to a generically optimized keyboard. Indeed, we found our personalization methods produced keyboards that—when evaluated amongst 16 participants without motor impairments against a generically optimized keyboard that was created using the same character transition occurrences—led to greater communication rates (via higher *ITR* and *WPM* values). Importantly, we did not find evidence of a speed–accuracy trade-off, as is often reported for virtual keyboard technology [80]; on average, participants improved in their ability to accurately select targets while increasing speed for personalized keyboards relative to the optimized keyboards. Although the keyboard demonstrated a significant, large main effect when comparing *ITR* and *WPM* across participants, a significant main effect was not demonstrated for target selection accuracy. This may be a byproduct of optimizing the keyboards for speed, not accuracy. Yet the benefits of increased speed provided by the keyboards also translate to *ITR*, a parameter that unifies both speed and accuracy. These results therefore highlight the importance of using metrics that unify movement time and selection accuracy as well as these individual components of alternative communication to comprehensively capture such benefits.

In addition to the observed communication improvements when using a personalized keyboard rather than a generically optimized keyboard, we observed significant relationships between communication outcomes and use time. Firstly, we observed that increased exposure to a keyboard led to improved communication via increased accuracy (albeit negligible effect size), *WPM*, and *ITR*. These findings imply that participants were able to communicate more effectively with increased familiarity with a keyboard interface. Our experimental paradigm was designed to create a new personalized keyboard for each participant after extensive use of the sEMG/IMU access method (~5 h) to test whether differences between keyboards persisted after gaining proficiency with the access method. Because there were significant differences in communication performance for the main effects of keyboard and block but *not* for the interaction of keyboard and block, our results indicate that participants gained familiarity with the access method over time and, further, that the personalized keyboards were superior for communicating even after learning the access method. Taken together, these results indicate that there is benefit in configuring characters based on transition probabilities *as well as* an individual's movement behaviors.

4.2.2. Personalized vs. QWERTY Keyboards

We hypothesized that a personalized keyboard would lead to better communication performance than QWERTY. Our results support this hypothesis, wherein keyboard (personalized, QWERTY) exhibited a significant main effect in the models for accuracy, *WPM*, *WPM\**, and *ITR*. Notably, the significant findings observed for *WPM\** indicates that differences in speed between keyboards cannot be attributed to the size and placement of the space key. Overall, these findings are of interest since participants reported high familiarity with QWERTY. Prior work suggests that it takes around 4–5 h of interaction to gain proficiency with an unfamiliar keyboard interface configuration [18]. Yet our results indicate that—even when factoring in additional visual search time required for an unfamiliar layout [18]—personalizing a keyboard to an individual's motor capabilities is an effective way to improve communication performance over using a QWERTY keyboard. These results highlight the previously regarded inefficiencies noted about QWERTY for single-input use [14,16,29].

4.3. *Limitations and Future Directions*

In this study, we developed and evaluated an automated method to personalize a virtual keyboard for AAC use. To minimize exposure to the ongoing COVID-19 pandemic in high-risk populations, we opted to examine the methodological feasibility in individuals without motor impairments rather than our intended target population of individuals with motor disabilities. It is thus unsurprising that a ceiling effect in target selection accuracy occurred across nearly all participants and all keyboards. Although outside the scope of this feasibility study, future work in control populations should examine the stability of outcome parameters (e.g., target selection accuracy) to account for differences in user control over time, as may occur in degenerative motor disorders such as amyotrophic lateral sclerosis. However, we examined Fitts' Law parameters at two timepoints (sessions 1 and 5) to generate personalized keyboards before and after sufficient exposure to the alternative access method. In addition to these results, however, our keyboard personalization methods successfully captured directional movement preferences amongst participants without motor disabilities. As such, this methodology shows promise for individuals with motor disabilities, especially for those with unequal ranges of movement or a visual field cut or condition that results in peripheral focus [34,35]. Future work therefore aims to employ similar methodology in the target population of people with motor disabilities.

This study was designed using a single-point access method configured on the forehead; we selected this configuration to minimize possible confounds by offering multiple access methods across multiple access points in establishing proof-of-concept effectiveness. With the diverse manifestation of neurological disorders, however, it is difficult to generalize the ability of our target population to sufficiently use the system with this fixed access method and, specifically, access point (i.e., forehead). By providing preliminary support for a directionally dynamic AAC system that can be personalized to an individual, future work will aim to expand access modalities. Due to our use of digraph transitions to facilitate key placement, the integration of additional keys such as numbers, 'return', and 'backspace', among other keys, was not possible for this current study, but with the expansion of the methods, these keys, as well as other functions offered by standard keyboards, could be employed in future work.

We additionally recognize the potential translation of this work to the field of human–computer interaction due to existing research that highlights performance variability with different access points across direction [36–39]. For example, given thumb performance differences with direction when using a mobile device [36], mobile device interfaces could be configured to minimize thumb flexion–extension movements compared to abduction–adduction movements. Additionally, expanding access methods alongside personalization methods such as SUPPLE [21] could yield interfaces beyond keyboards that are flexible for people with or without motor impairments. Within this vein, we acknowledge that our system could integrate common communication options offered in many AAC devices—such as word completion, dynamic target size, and dwell click—and think these options would also be valuable to include as we expand the capabilities of our system.

## 5. Conclusions

In this work, we present and evaluate methods to develop personalized, virtual keyboard interfaces for alternative communication against existing methods for computationally optimizing keyboards as well as the standard QWERTY keyboard. Communication performance benefits were observed when using a personalized keyboard compared to existing optimized keyboards as well as the QWERTY keyboard. Our results suggest that the benefits provided by personalized keyboards are related to the combined improvements in the speed and accuracy of selecting characters on the keyboard to construct messages. Overall, our results show that personalizing a keyboard is an effective strategy to improve communication with a single-input AAC access method and can achieve communication rates higher than the ubiquitous QWERTY layout. This work highlights the first instance of applying automated algorithms to tailor the layout of a virtual keyboard to an individual based on their motor abilities and, moreover, demonstrate promise for using ability-based methods to design personalized assistive technology.

**Author Contributions:** conceptualization, C.L.M., G.J.C., S.K.F., J.C.K. and J.M.V.; methodology, C.L.M., J.C.K. and J.M.V.; software, C.L.M.; validation, C.L.M. and J.M.V.; formal analysis, C.L.M. and J.M.V.; investigation, C.L.M. and J.M.V.; resources, G.J.C., S.K.F., P.C., S.H.R., G.D.L., J.C.K. and J.M.V.; data curation, C.L.M. and J.M.V.; writing—original draft preparation, C.L.M.; writing—review and editing, C.L.M., G.J.C., S.K.F., P.C., S.H.R., G.D.L., J.C.K. and J.M.V.; visualization, C.L.M.; supervision, J.C.K. and J.M.V.; project administration, J.M.V.; funding acquisition, P.C. and J.M.V. All authors have read and agreed to the published version of the manuscript.

**Funding:** This research was funded by the National Institute on Deafness and Other Communication Disorders of the National Institutes of Health under award number R43DC018437 and by the De Luca Foundation.

**Institutional Review Board Statement:** The study was conducted in accordance with the Declaration of Helsinki and approved by the Western Institutional Review Board (WIRB® Protocol #20192468, approved 23 September 2021).

**Informed Consent Statement:** Informed consent was obtained from all subjects involved in the study.

**Data Availability Statement:** The data underlying the results presented in the study are available from Altec, Inc. at contact@altecresearch.com.

**Conflicts of Interest:** C.L.M., P.C., S.H.R., G.DL., J.C.K. and J.M.V. are employed by Delsys, a private company that manufactures, markets, and sells electromyographic sensors and other physiological measurement systems.